\documentclass[preprint]{elsarticle}

\usepackage{subfigure}

\newcommand\vv{\mathbf v}
\newcommand\ud{\mathrm d}
\newcommand\ut{\hat{\mathbf t}}

\title{Observation of a non-adiabatic geometric phase for elastic waves}
\author{J\'er\'emie~Boulanger}
\author{Nicolas Le Bihan}
\address{Université Grenoble 4 / CNRS,\\
         Gipsa-lab -
         BP 46, 38402 Saint-Martin d'H\`eres, FRANCE}
\author{Stefan Catheline}
\address{Universit\'e Joseph Fourier - Grenoble 1 / CNRS, \\
         Institut des Sciences de la Terre -
         BP 53, 38041 Grenoble, FRANCE}
\author{Vincent Rossetto\corref{vincent}}
\address{Universit\'e Joseph Fourier - Grenoble 1 / CNRS,  \\
         Laboratoire de Physique et Mod\'elisation des Milieux Condens\'es -\\
         BP 166, 38042 Grenoble, FRANCE}
\cortext[vincent]{Corresponding author
\ead{vincent.rossetto@grenoble.cnrs.fr}}

\begin{document}
\begin{abstract}
We report the experimental observation of a geometric phase for elastic waves
in a waveguide with helical shape. The setup reproduces the experiment by
Tomita and Chiao ({\it Phys. Rev. Lett.} {\bf 57}, 1986) that showed first
evidence of a Berry phase, a geometric phase for adiabatic time evolution, in
optics.  Experimental evidence of non-adiabatic geometric has been reported in
quantum mechanics.  We have performed an experiment to observe the polarization
transport of classical elastic waves. In a waveguide, these waves are polarized
and dispersive.  Whereas the wavelength is of the same order of magnitude as
the helix' radius, no frequency dependent correction is necessary to account
for the theoretical prediction.  This shows that in this regime, the geometric
phase results directly from geometry and not from a correction to an adiabatic
phase.  
\end{abstract}

\maketitle

\section{Introduction}
\label{sec:intro}
Polarization is a feature shared by several kinds of waves:
light and elastic waves, for instance, have two transverse polarization
modes \cite{bornwolf}. The
polarization degrees of freedom are constrained to lie in the plane
orthogonal to the direction of propagation. This constraint is responsible,
in optics, for the existence of a geometric phase. Geometric phases
of different kinds have been discovered after the Berry phase~%
\cite{berry1984,hannay1985,aharonov1987,anandan1988,zwanziger1990,shaperewilczek,bhandari1997}.

The geometric phase of light was first experimentally observed
by Tomita and Chiao \cite{tomita1986} using optical fibers.
It was later suggested that a geometric phase should exist for 
any polarized waves \cite{segert1987}. This Letter discusses
the case of elastic waves when the adiabatic conditions are not fulfilled,
a situation which can not be reached in Optics.
For polarized waves in a waveguide, the geometric phase differs
from zero only if the shape of the waveguide is three-dimensional
and is defined even if the time evolution is not cyclic \cite{samuel1988}. 

The fundamental origin of geometric phases lies in the 
geometric description of the phase space. The existence of a
geometric phase is related to the curvature, either local or global,
of the phase space.
In a first attempt to classify geometric phases, Zwanziger~{\it et al.}
distinguish adiabatic geometric phases~\cite{zwanziger1990}, 
the main example of which is a spin in a magnetic field \cite{berry1984}.
The geometric phase for the spin is
defined if the system evolves adiabatically, such that transitions
between spin states are negligible. The direction of the magnetic field
must therefore evolve at a rate~$1/T$ much smaller than the oscillation
frequency between spin eigenstates. 

Consider the case of waves propagating in a curved waveguide.
The role of the magnetic
field's direction is played by the direction of propagation
and the phase between the spin eigenstates is the orientation of
the linear polarization of the wave.
The adiabatic approximation imposes that
the evolution rate~$1/T$ is much smaller than the wave frequency.
In the first experimental evidence for the adiabatic geometric phase,
performed in optics by Tomita and Chiao \cite{tomita1986},
the frequency of light~$\nu$ was indeed several orders of magnitude larger
than~$1/T$. Photon spin flip is negligible, therefore the adiabaticity
conditions are fulfilled. 
In Foucault's pendulum \cite{vonbergmann2007}, a renowned case of 
classical geometric phase, these conditions are fulfilled as well.

Some geometric phases do not require
adiabaticity, such as Pancharatnam phase \cite{pancharatnam1956} and the
Aharonov-Anandan quantum phase \cite{aharonov1987} or certain canonical
classical angles \cite{anandan1988,berry1988}.
Geometric phases have been observed
in many fields of science and called with different names. In classical
mechanics, the geometric phase for adiabatic invariants is often referred to as
\emph{Hannay angle} \cite{hannay1985,berry1996}, in knot theory and DNA
physics, the name~\emph{writhe} is mostly used \cite{agarwal2002,rossetto2003}.

We consider from now on elastic waves, which
polarization state can be represented as a combination 
of two linearly polarized states. These are classical waves, quantum
transition between polarization eigenstates is not possible. 
The adiabaticity condition
$\nu\gg 1/T$ should therefore not be required to observe a 
geometric phase. In our experiment, indeed, the frequency and the evolution
rate have the same order of magnitude. 
In this Letter, we briefly introduce the geometric phase
in a purely geometric picture and compute its value along a helix.
We present an experiment designed to
measure the geometric phase of elastic waves in a helical 
waveguide with the condition $\nu\simeq 1/T$. 
Without loss of generality, we only consider linear 
polarization.

Although the experiment we investigate has many common points with previous
studies, some distinctions must be pointed out. Contrary to light polarization,
the phase cannot be interpreted as a quantum phase difference between two
eigenstates.
There is no established classification of geometric phase but as the
system we study is purely classical and neither cyclic nor adiabatic,
the observed geometric phase cannot be rigorously identified as a Berry, 
Pancharatnam, Aharonov-Anandan or Wilczek-Zee phase, for instance.
The classical geometric angles \cite{anandan1988,berry1988} follow from
the action-angle representation of the system, which is valid for the
motion of a material point of the waveguide, but does not rigorously
apply to the elastic wave transport. 

\section{The geometric phase for polarized waves}
\begin{figure}
  \begin{center}
    \subfigure{\includegraphics[width=0.4\textwidth]{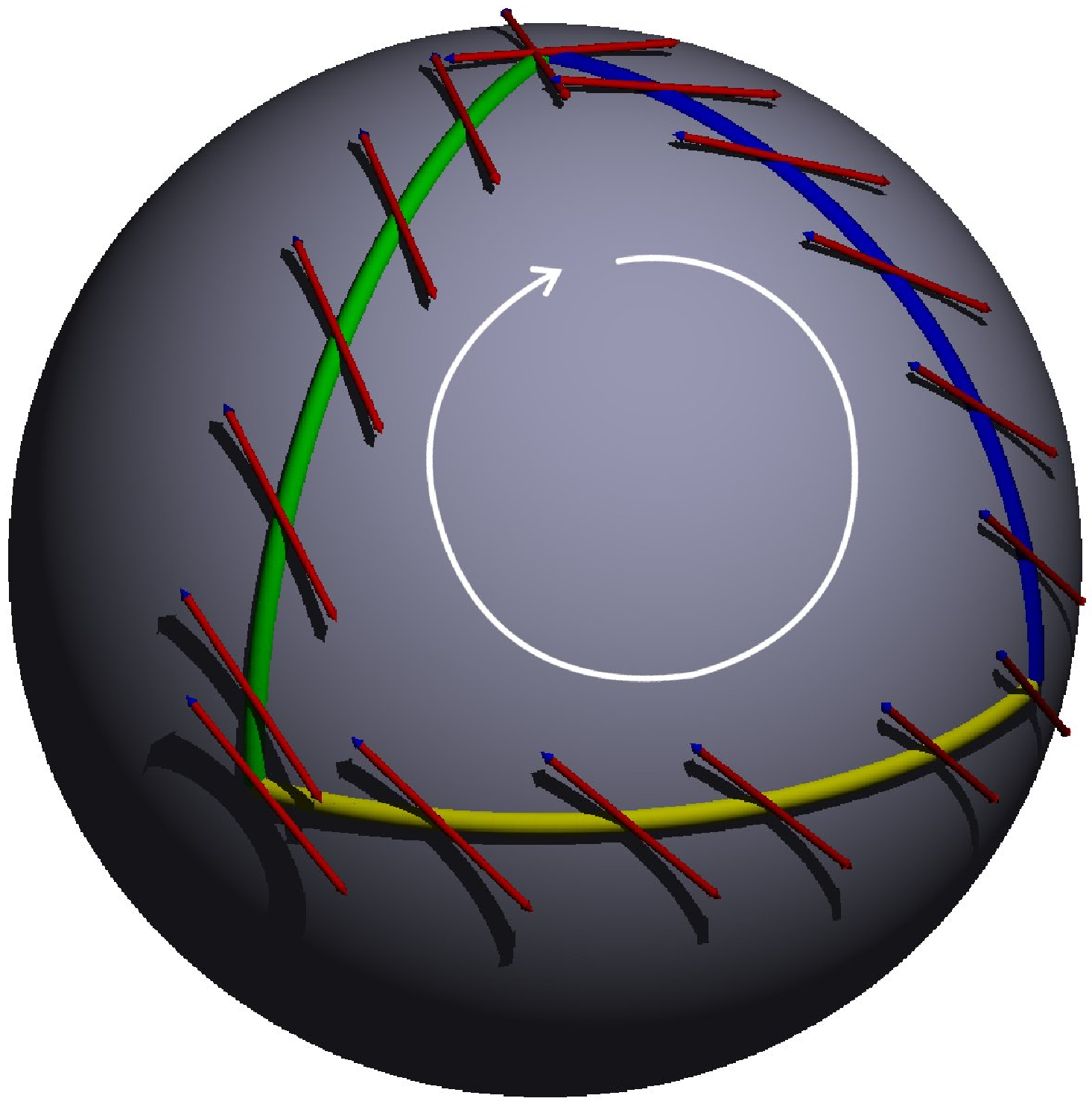}}
    \subfigure{\includegraphics[width=0.3\textwidth]{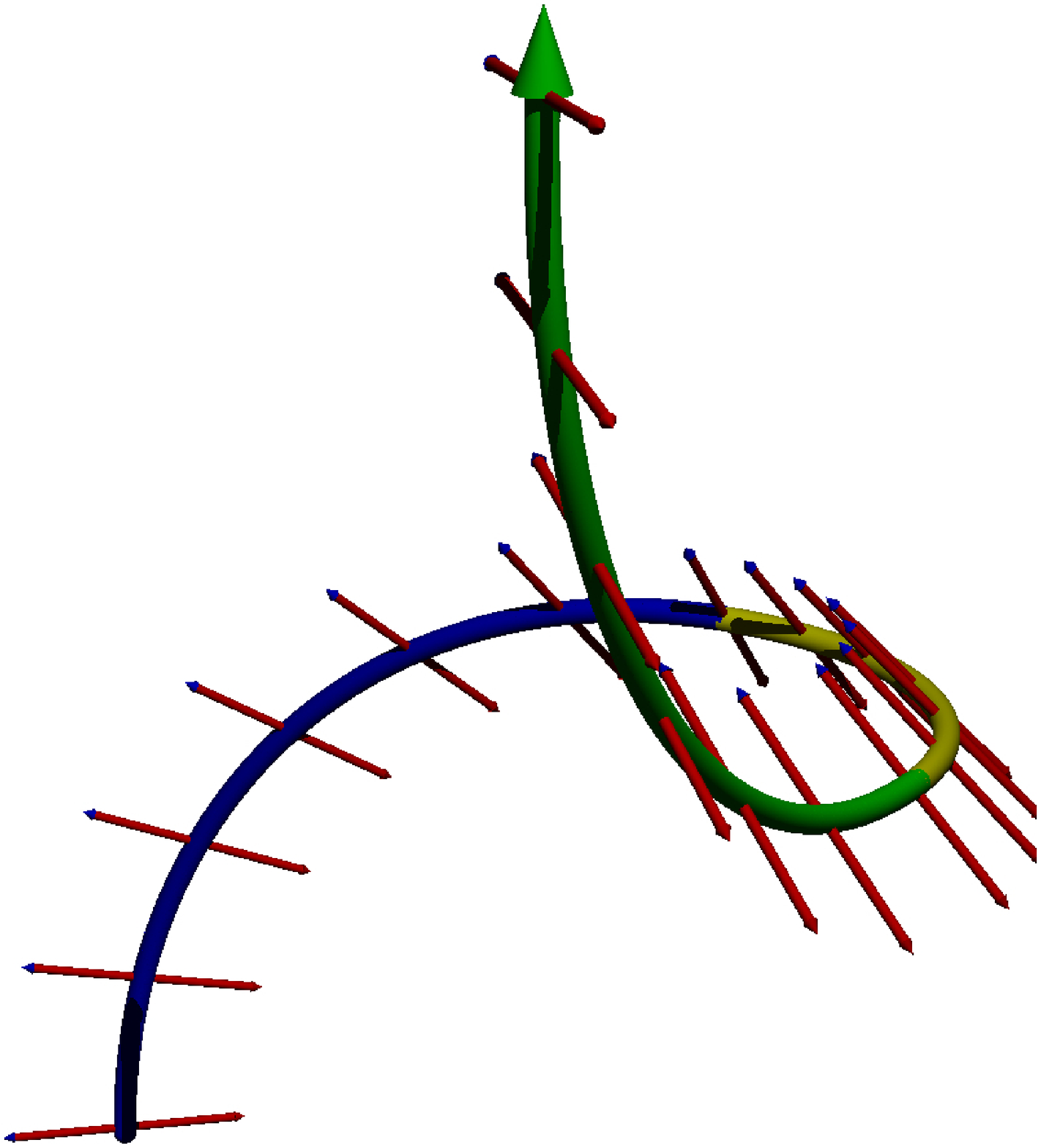}}
    \caption{\label{fig:loop} 
    (Left) A trajectory represented on the sphere of tangent vector.
    Linear polarization is represented
    as a direction in the plane orthogonal to the direction of propagation
    and the trajectory is made
    of three arcs of great circles forming a right angle with each
    others. Both at the start and the end of
    the trajectory (upper pole), the tangent vector points upwards.
    One observes the anholonomy of parallel transport:
    The double arrow indicating the polarization direction, 
    although it is parallel transported all along the trajectory
    back to its initial position, has different directions at 
    the beginning and at the end of the displacement. 
    (Right) Same trajectory in real space.} \end{center}
\end{figure}

A wave travelling along a straight path keeps a constant polarization 
along the trajectory; if the direction of propagation is not constant,
as polarization is ascribed to remain in the orthogonal plane,
it evolves along the path.
The path-dependent transformation
transporting polarization must be, for 
physical reasons, linear, reversible and continuous. There is 
only one transformation satisfying these requirements:~%
\emph{parallel transport}~\cite{segert1987}.
Along the path followed by the wave, the direction
of propagation is represented as a point on the unit sphere,
polarization is represented in the tangent plane to the sphere 
and transported in the sphere's tangent bundle, see Figure~\ref{fig:loop}.

Let us consider a geodesic on the unit sphere, {\em i.e.} an arc 
of a great circle.
If polarization is colinear or orthogonal to the great circle, 
it remains so along the geodesic to preserve symmetry. 
Any polarization is a linear combination of these two particular
linear polarizations. The linearity of parallel transport implies, for linear
polarizations, that if one rotates the initial polarization
in the initial tangent plane, the parallel transported polarizations
rotate by the same angle in all tangent planes of the tangent bundle.
Parallel transport can be extended to any smooth and piecewise
differentiable trajectory of the tangent vector on the unit sphere
by discretizing the path into elementary arcs of great circles~%
\cite{rossetto2002}.

A vector parallel transported along a closed trajectory
does not necessarily have the same orientation in the tangent plane
at the starting point after one stride along the trajectory. The angle
between the initial and final vectors is algebraically equal to the area
enclosed by the trajectory on the sphere \cite{nakahara}. In the
Figure~\ref{fig:loop}, this area is equal to~$\pi/2$. It is also known
as the anholonomy of the trajectory in the phase space.

\section{Computation of the phase}
All the distances, denoted by $s$, are given in arclength along the
helix and the accelerometers position is taken as the origin. We call~$R$ the radius of the helix and~$P$ its pitch, 
and define~$L=\sqrt{(2\pi R)^2+P^2}$. The angle~$\theta$ between the
direction of propagation~$\ut(s)$
and the helix main axis is given by~$\sin\theta=P/L$. The Fr\'enet 
torsion~$\tau=(2\pi\cos\theta)/L$ of the helix
is constant. $2\pi s/L$ is the azimuth angle spanned by the tangent 
vector along a distance $s$ on the helix. 
Fuller's theorem states that the geometric phase difference between two
trajectories is equal to the area spanned by the tangent vector
on the unit sphere during one smooth deformation of one trajectory
into the other. 
We chose as reference the trajectory $\theta=\pi/2$ 
(the limit case where the helix is a flat circle) for which it is
known that the geometric phase vanishes because the
trajectory is two-dimensional. After deformation of the reference into
the helix, Fuller's theorem yields 
a phase proportional to $2\pi s/L$ 
and the proportionality coefficient is $\cos\theta$ (given by the classic geometry formula for the spherical cap area). 
We obtain a geometric phase $\Phi(s)$ of:

\begin{equation}
\Phi(s)=(2\pi s/L) (\cos\theta) = \tau s.
\end{equation}

From an intrinsic point of view, parallel transport corresponds to 
transporting a polarization vector~$\vv$ 
while keeping it constant for an imaginary walker striding
along the trajectory. In the language of differential geometry
the vector's covariant derivative must equal~zero
along the trajectory~\cite{nakahara}. 
We obtain the equation of parallel transport:
\begin{equation}
D\vv=\ud\vv-(\ud\vv\cdot\ut)\ut=0.
\end{equation}
The components of~$\vv$
in the Fr\'enet-Serret frame follow a differential equation
describing a rotation at the rate~$\tau$, leading to a
phase:
\begin{equation}
\Phi(s)=\tau s.
\end{equation} 

\section{Experiment}
\begin{figure}
\begin{center}
  \includegraphics[width=0.4\textwidth]{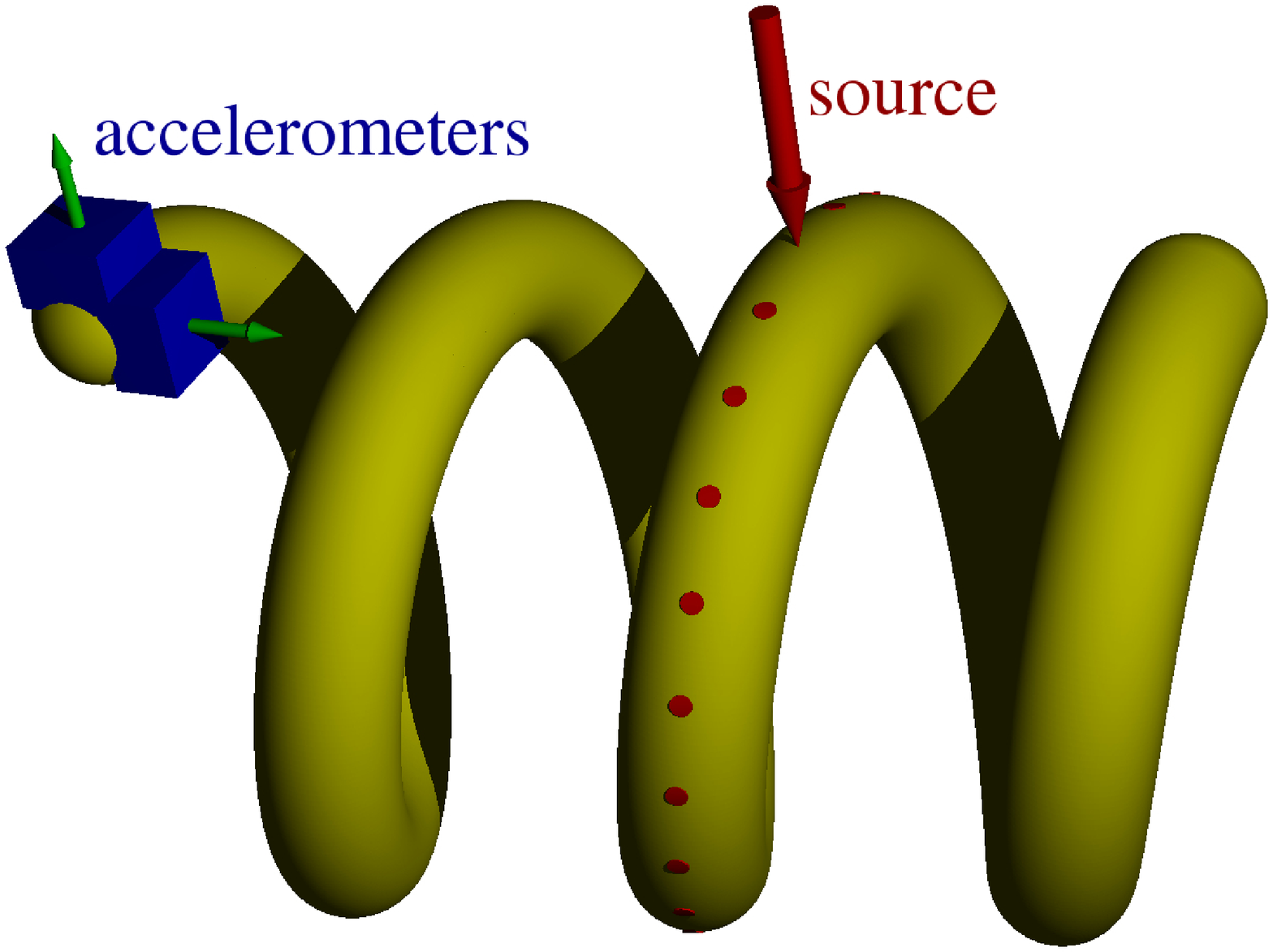}
  \includegraphics[width=0.4\textwidth]{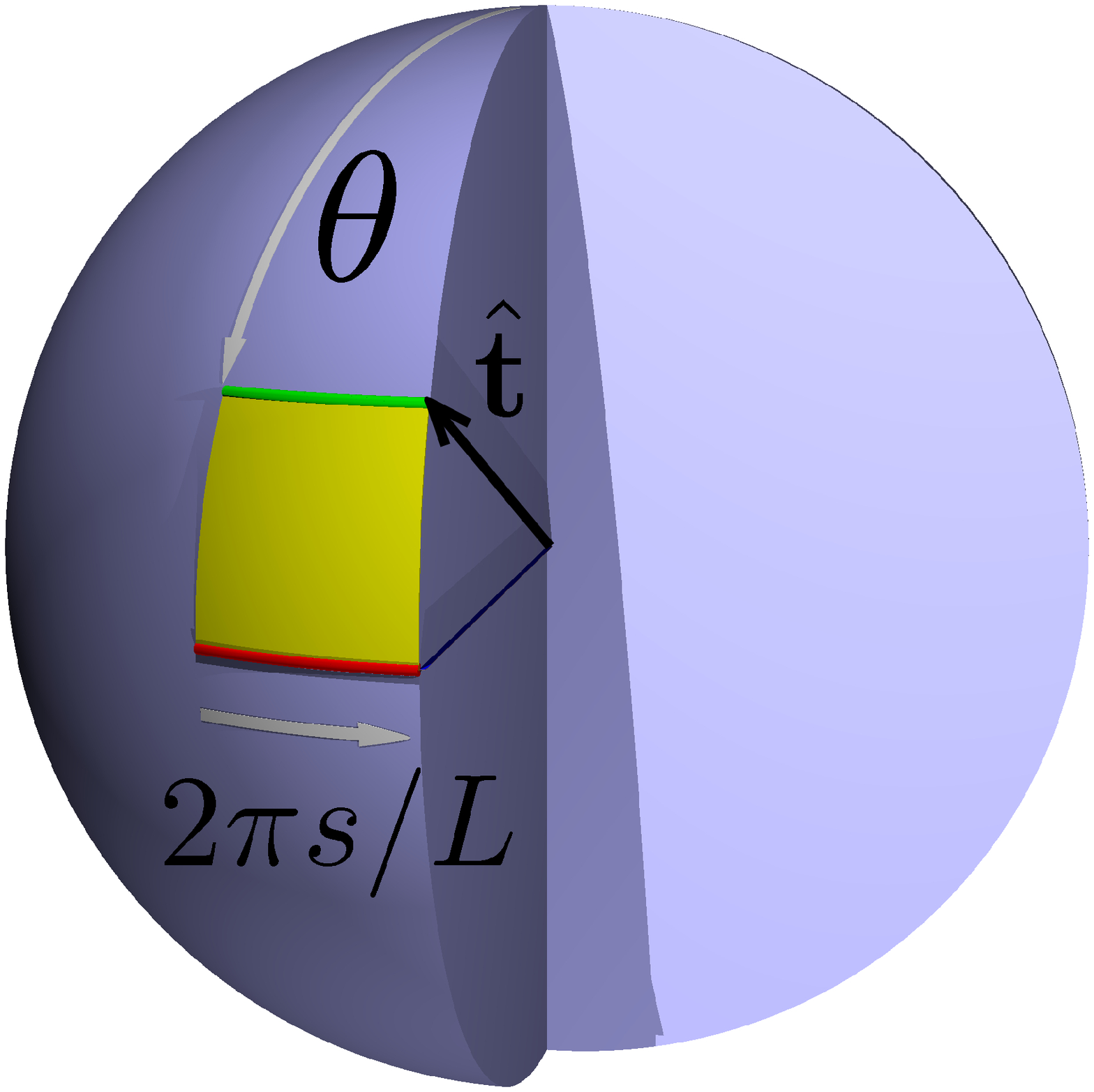}
  \caption{\label{fig:manip} 
  (Left) Setup of the experiment. 
  Accelerometers are represented as cubes, 
  they record acceleration in two orthogonal directions (small arrows).
  The large arrow symbolizes the source, a radial impact on the helix. 
  The dots indicate some of the source positions.
  (Right) Geometry of the transformation from
    the reference trajectory to the helix trajectory. 
    During the transformation the tangent vector for~$s$ fixed 
    follows a meridian (vertical arrow on the figure)
    between colatitudes~$\pi/2$ and~$\theta$.
    A displacement of $s$ corresponds to an azimuth angle of
    $2\pi s/L$ (horizontal arrow). The area 
    spanned by the tangent vector during the deformation is equal to the 
    geometric phase difference computed by Fuller's formula. }
  \end{center}
\end{figure}

\begin{center}
\begin{figure}
\begin{center}
\subfigure{\includegraphics[width=0.44\textwidth]{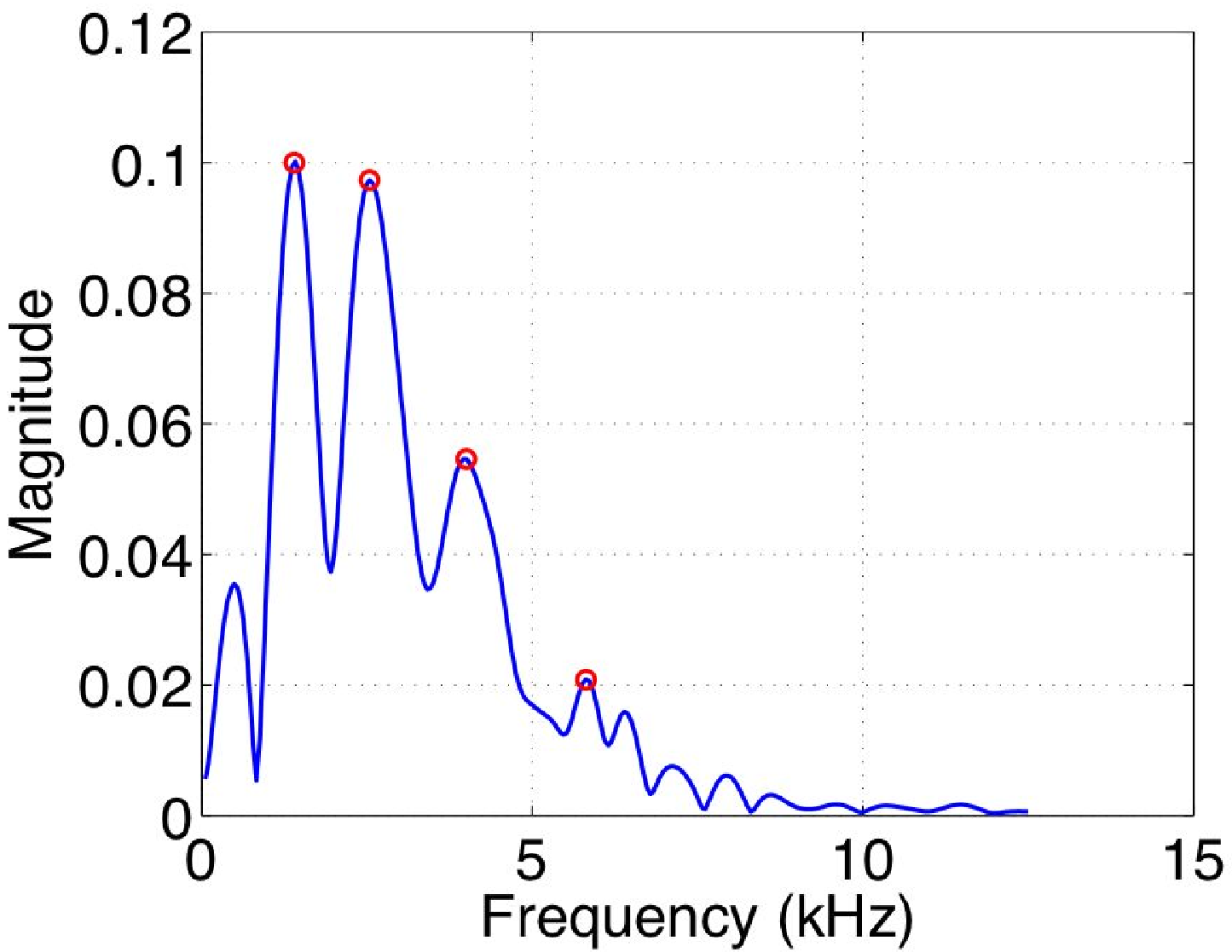}}
\subfigure{\includegraphics[width=0.44\textwidth]{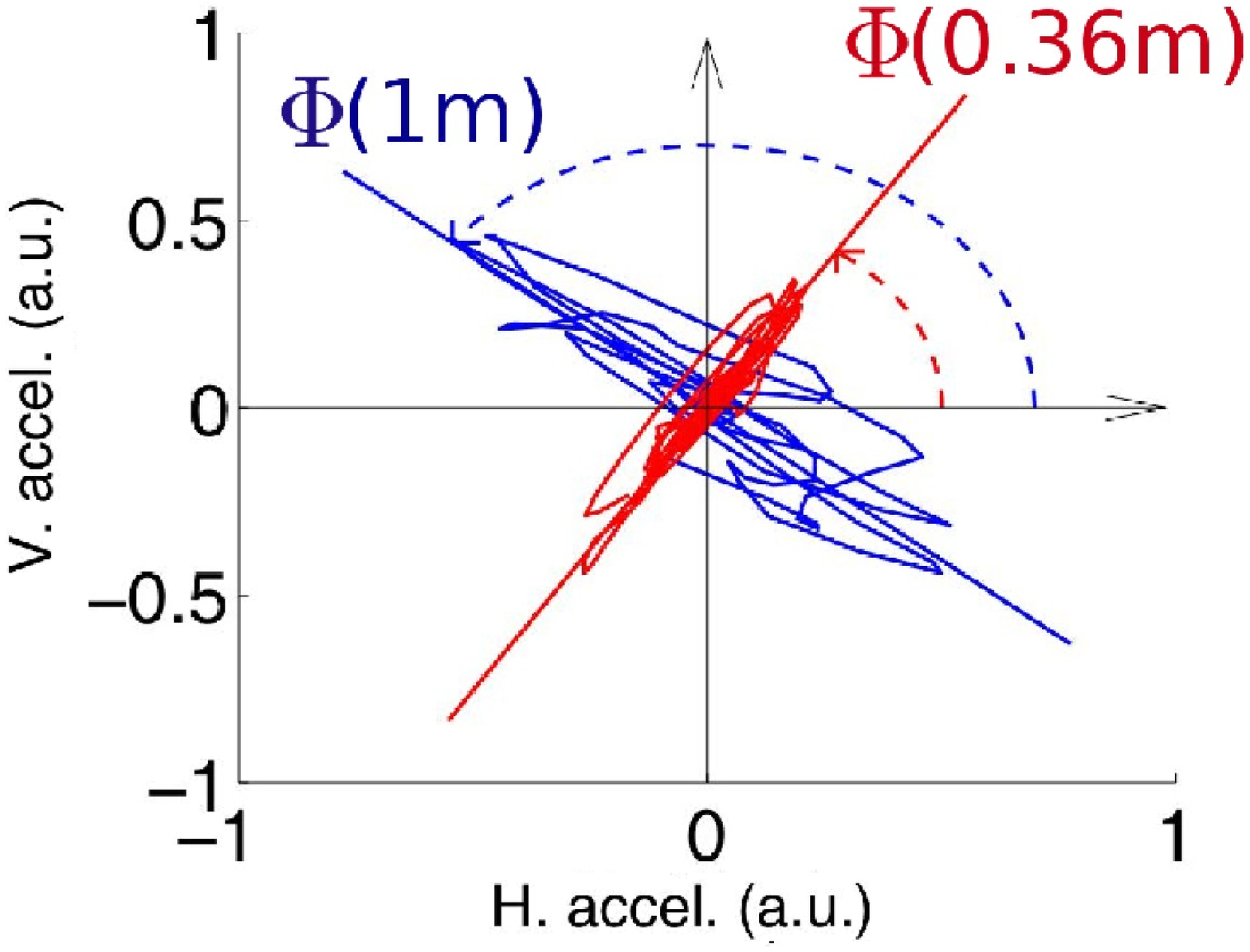}}
\caption{\label{fig:Waveform}(Left) Power spectrum of the first
  oscillations of the signal corresponding to direct arrival of the
  flexural mode. Circles indicate the maxima used as central
  filtering frequencies in the signal analysis.  (Right) Parametric
  plot of the first oscillations (direct arrival) for sources located
  at $s=0.36\,\mathrm{m}$ and $s=1.00\,\mathrm{m}$. The relative angle
between polarization orientations is the geometric phase difference.}
\end{center}
\end{figure}
\end{center}

In order to reproduce the setup used by Tomita and Chiao~\cite{tomita1986},
we use a metallic spring as a waveguide for elastic waves,
taken from a car's rear damper. 
It has a circular section of $r=13.5\,\mathrm{mm}$ making
five coils of radius $R=75\pm1\,\mathrm{mm}$ and with a pitch of 
$P=91.5\pm1\,\mathrm{mm}$. The direction of propagation makes a constant
angle~$\theta=1.379\,\mathrm{rad}$ with the helix' main axis.
As the cross-section of the helix is circular, the 
flexural modes are degenerated.
The helix is
suspended to two strings to isolate the system.
We use two accelerometers (Br\"uel \& Kjaer, type 4518-003)
located at one end of the spring and record vibrations in two orthogonal directions
(see Figure~\ref{fig:manip}).
The sampling frequency being set to $50\,\mathrm{kHz}$,
the information in the signal is available up to $25\,\mathrm{kHz}$. 
Considering the accelerometers' power spectrum 
in their stability range~$1\mathrm{kHz}-25\mathrm{kHz}$ 
(see Figure~\ref{fig:Waveform}),
frequencies above $5\,\mathrm{kHz}$ can be ignored.
The signal is amplified (Br\"uel \& Kjaer amplifier,
type 2694) before signal processing. 

\begin{figure}
\begin{center}
  \includegraphics[width=0.9\textwidth]{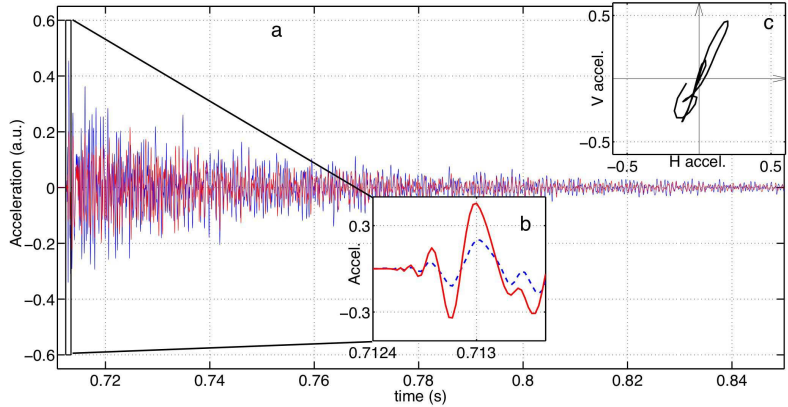}
  \caption{\label{fig:signaux} Example of vertical (solid line) and
    horizontal (dashed line) displacement signals recorded during the
    experiment. (a): complete recorded waveforms associated to one
    source. (b): time windowed signals considered
  for polarization orientation estimation. (c): associated
  polarization parametric plot. The truncated (time windowed) signal is chosen in
  order to isolate direct linearly polarized wave from other modes and
reflected waves. Only few first oscillations are considered.}
\end{center}
\end{figure}

We record the waves generated by 32 equally spaced sources, which distance from the accelerometers 
ranges from~$s=5\,\mathrm{cm}$ to $s=1.45\,\mathrm{m}$. 
Making weak impacts on the metal spring creates bending 
waves that are linearly
polarized. Impacts are made radially with respect to the helix. 
The source signal is generated manually,
by gently hitting the waveguide with a hammer at the different source
positions. Polarization depending only on the amplitude ratio measured
by the accelerometers, it is not sensitive to the energy of the source.

The wave propagation modes in helical waveguide constitute a difficult
problem. Solutions can only be found numerically \cite{treyssede2007}
and yield a complex mode structure. The fundamental mode is a flexural
mode with linear polarization. Polarization losses appear rapidely
after the first waveforms. In order to measure the geometric phase, we use 
only the first periods of oscillation of the signal, before depolarization
is complete. In figure \ref{fig:signaux}, the time windowed signals
used for polarization orientation estimation are presented, together
with the complete waveforms and associated polarization parametric
plots. In Figure \ref{fig:Waveform}, we show 
an example of two parametric polarization plots for signals 
obtained with the source at different distances from the
accelerometers. The relative angle between polarization orientations
is the geometric phase difference.

Bending waves in the waveguide are dispersive. We therefore filtered
the signal using non-overlapping bandpass filters centered at frequencies:
$1.4\,\mathrm{kHz}$, $2.5\,\mathrm{kHz}$, $4.0\,\mathrm{kHz}$ and
$5.8\,\mathrm{kHz}$. These values correspond
to maxima of the power spectrum displayed in figure \ref{fig:Waveform}
(left). The corresponding velocities are displayed in the Table~\ref{table}. 

The polarization orientation is obtained from the
records using a principal component analysis~\cite{Vidale1986}.
This technique consists in obtaining the eigenvectors of the cross-correlation
between the two orthogonal signals (recorded on the two accelerometers). The eigenvector with the largest eigenvalue
gives the polarization direction. The geometric phase~$\Phi(s)$ is thus the 
orientation of polarization measured when the source is located at distance~$s$
from the accelerometers (see Figure~\ref{fig:manip}).

\begin{figure}
\begin{center}
\includegraphics[width=0.9\textwidth]{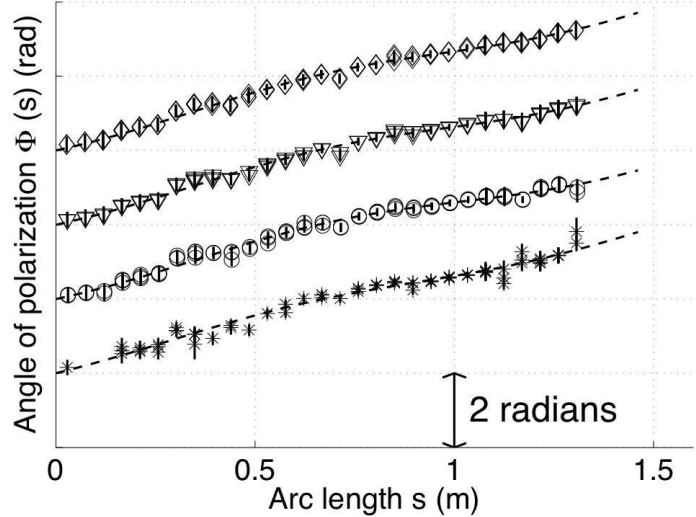}
\end{center}
\caption{\label{fig:phasegeo5}
Measured geometric phase as a function of arc length along the helix 
for signals filtered at $1.4\,\mathrm{kHz}$, $2.5\,\mathrm{kHz}$,
$4.0\,\mathrm{kHz}$ and $5.8\,\mathrm{kHz}$.
The dashed 
line represents the linear regression with correction
for the coupling and orthogonality of the accelerometers.
The origins of the curves are arbitrary.}
\end{figure}

We perform several measurements for each position of the source, in
order to reduce effects of external perturbations and variations of
the source signal. 
In our results, we observe oscillations that are due to
imperfections related to the accelerometers response and setup.
Denoting by $\epsilon$ the shift from orthogonality
between the accelerometers and~$1+\eta$ the ratio 
of their mechanical couplings, the principal components analysis
yields:
\begin{equation}
\Phi(s)=\Phi(0)+
     \tau s+\sin\Phi(s)\big(\eta\cos\Phi(s)-\epsilon\sin\Phi(s)\big).
\end{equation} 
We obtain a linear coefficient~$\tau\simeq2.5\;\mathrm{rad.m^{-1}}$ 
for the signals filtered at four differents frequencies
(bandwidth~$\pm20\,\mathrm{\%}$).
Numerical values are presented in Table~\ref{table}. 

\begin{table}
  \begin{center}
  \begin{tabular}{|c|c|c|c|c|c|}
    \hline
    $\nu (\mathrm{kHz})$
    & $c (\mathrm{m.s^{-1}})$
    & wavelength ($\mathrm{m}$)
    & $\tau$ ($\mathrm{rad.m^{-1}}$) 
    & $\eta$ 
    & $\epsilon$ ($\mathrm{rad}$) \\
    \hline
    $1.4$ & $2 \,10^3$ & $1.4$ & 2.53 & 0.31 & -0.03 \\
    $2.5$ & $3 \,10^3$ & $1.2$ & 2.35 & 0.45 &  0.03 \\
    $4.0$ & $4 \,10^3$ & $1.0$ & 2.45 & 0.39 &  0.05 \\
    $5.8$ & $3.5 \,10^3$ & $0.6$ & 2.50 & 0.38 &  0.07 \\
    \hline
  \end{tabular}
  \caption{\label{table}Frequencies, velocities, wavelengths and
   fit parameters for the filtered signals. 
   The values of~$\tau$ have a confidence range 
   of~$\pm0.1\,\mathrm{rad.m^{-1}}$.}
  \end{center}
\end{table}

\section{Results and discussion}
The theoretical Fr\'enet torsion of the helix used in the experiment is 
$\tau=2.49\pm0.1\,\mathrm{rad.m^{-1}}$.
The fits performed from experimental data give
an estimation of $\tau=2.50\pm0.1\,\mathrm{rad.m^{-1}}$. 
We do not observe a significant dependency of~$\tau$ with
respect to the frequency or the velocity (see Table \ref{table}), which justifies 
the denomination ``geometric phase''~: the effect we observe is
solely due to the geometry of the waveguide.

Apart from bending waves, compression waves and
torsion waves can also propagate in the helix.
In a straight rod, these propagation modes are not coupled
and bending waves remain polarized at long times.
In helices, the coupling between
compression waves and torsion waves increases
with curvature and torsion.
Bending waves and torsional waves are therefore partially converted into
each other, back and forth, during the propagation.
This explains depolarization in
our measurements and why we consider only the first 
oscillations of the signals.

We have filtered the polarized waves at four frequencies
of the order of magnitude of the rate at which the propagation
direction evolves $1/T=c/L$.
Such evolution rates imply that the regime is not adiabatic. 
Thanks to the dispersivity of the bending waves in the system,
the adiabatic parameter~$c/L$ can be varied by changing the filtering
central frequency.
Therefore filtering at several frequencies is equivalent
to explore the non-adiabatic regime.
No significant variations of the results are observable in the 
frequency range, which is an experimental evidence of the non-adiabaticity 
of the geometric phase, complementary to the theoretical derivation.
Converted into lengths, this means that the wavelength of the bending
wave is of the same order of magnitude as the length of a helix coil.

One may argue that the conversions between modes observed in the
experiment are the classical equivalent of the transition between
spin eigenstates of the quantum problem. By nature, however, these
transitions are different. In the quantum problem, the transitions
are of the first order in time but for the elastic waves, depolarization 
processes are of the second order, because an intermediate, unpolarized
state is necessary. According to the dispersion relation in a helical
waveguide~\cite{treyssede2007}, the intermediate state is mainly the 
torsional mode, and second the longitudinal compression mode.

\section{Conclusion}
We have demonstrated the existence of a geometric phase for
elastic waves in a waveguide far from the adiabatic regime.
We have investigated the results at several frequencies,
or equivalently with several values of the adiabatic parameter,
and observed no significant variations of the phase's value.
This is an experimental evidence, in addition to the theoretical
derivation, that the
observed phase is then not an adiabatic geometric phase
with non-adiabatic corrections, but a non-adiabatic geometric phase
for classical systems.
Because there is no need to invoke the adiabatic approximation to
preserve the polarization state, the 
geometric phase should extend to all frequencies in 
smooth waveguides with circular sections, a interesting result for waves 
with very large wavelengths.
In quantum mechanics, the Aharonov-Anandan phase~\cite{aharonov1987} 
shares the same non-adiabatic aspect as the non-quantum geometric
phase studied in this letter.

In nature, polarized elastic waves, such as seismic S~waves (shear waves)
are observed under certain conditions, and the concept of geometric
phase applies. The degree of polarization and the statistics of
the polarization orientation 
in seismic signals created by a polarized source contain
informations concerning the disorder of propagating medium
that can be understood in terms of geometric phases.

\section{Acknowledgments}
The authors would like to thank \'E. Larose for discussions
and B. de Cacqueray for his experimental help.
This work was partially funded by the ANR-JC08-313906
SISDIF and CNRS/PEPS-PTI grants.

\bibliography{phase_elastic}
\end{document}